\pdfoutput=1
\documentclass[a4paper]{jpconf}
\usepackage{graphicx}
\usepackage{epstopdf}
\usepackage{multirow,dcolumn}
\usepackage{geometry}
\usepackage{color}

\begin{document}
\newcommand{\nnprime}{n,n$^\prime \gamma$}
\newcommand{\ntwon}{n,2n$\gamma$}
\newcommand{\nthreen}{n,3n$\gamma$}
\newcommand{\nxn}{n,xn$\gamma$}
\newcommand{\nx}{n,x$\gamma$}

\newcommand{\natpb}{$^{\textrm{nat}}$Pb}
\newcommand{\natge}{$^{\textrm{nat}}$Ge}
\newcommand{\eigpb}{$^{208}$Pb}
\newcommand{\sevpb}{$^{207}$Pb}
\newcommand{\sixpb}{$^{206}$Pb}
\newcommand{\fivpb}{$^{205}$Pb}
\newcommand{\foupb}{$^{204}$Pb}
\newcommand{\nonubb}  {$0 \nu \beta \beta$}
\newcommand{\twonubb} {$2 \nu \beta \beta$}
\newcommand{\gam}{$\gamma$}
\def\nuc#1#2{${}^{#1}$#2}
\def\mee{$\langle m_{\beta\beta} \rangle$}
\def\mnu{$\langle m_{\nu} \rangle$}
\def\ml{$m_{lightest}$}
\def\gnu{$\langle g_{\nu,\chi}\rangle$}
\def\mmod{$\| \langle m_{\beta\beta} \rangle \|$}
\def\mb{$\langle m_{\beta} \rangle$}
\def\BBz{$0 \nu \beta \beta$}
\def\BBm{$\beta\beta(0\nu,\chi)$}
\def\BBt{$2 \nu \beta \beta$}
\def\nonubb{$0 \nu \beta \beta$}
\def\twonubb{$2 \nu \beta \beta$}
\def\BB{$\beta\beta$}
\def\Mz{$M_{0\nu}$}
\def\Mt{$M_{2\nu}$}
\def\MzG{$M^{GT}_{0\nu}$}           
\def\MzF{$M^{F}_{0\nu}$}                
\def\MtG{$M^{GT}_{2\nu}$}           
\def\MtF{$M^{F}_{2\nu}$}                
\def\Tz{$T^{0\nu}_{1/2}$}
\def\Tt{$T^{2\nu}_{1/2}$}
\def\Tc{$T^{0\nu\,\chi}_{1/2}$}
\def\Rz{$\Gamma_{0\nu}$}            
\def\Rt{$\Gamma_{2\nu}$}            
\def\ms{$\delta m_{\rm sol}^{2}$}
\def\ma{$\delta m_{\rm atm}^{2}$}
\def\ts{$\theta_{\rm sol}$}
\def\ta{$\theta_{\rm atm}$}
\def\tot{$\theta_{13}$}
\def\gpp{$g_{pp}$}                  
\def\qval{$Q_{\beta\beta}$}                 
\def\MJ{{\sc Majorana}}             
\def\DEM{{\sc Demonstrator}}             
\def\be{\begin{equation}}
\def\ee{\end{equation}}
\def\cpRty{counts/ROI/t-y}
\def\onecpRty{1~count/ROI/t-y}
\def\fourcpRty{4~counts/ROI/t-y}
\def\ppc{P-PC}                          
\def\nsc{N-SC}                          
\def\instLANL{$^1$}			
\def\instPNNL{$^2$}		
\def\instUNC{$^{3}$}	
\def\instTUNL{$^{4}$}	
\def\instLBNLed{$^{5}$}		
\def\instUW{$^{6}$}	
\def\instORNL{$^7$}
\def\instUSC{$^8$}
\def\instNCSU{$^{9}$}	
\def\instITEP{$^{10}$}
\def\instUC{$^{11}$}		
\def\instLBNLnsd{$^{12}$}		
\def\instJINR{$^{13}$}
\def\instUT{$^{14}$}		
\def\instOU{$^{15}$}		
\def\instDUKE{$^{16}$}	
\def\instUA{$^{17}$}		
\def\instLBNLinpa{$^{18}$}		
\def\instUCB{$^{19}$}	
\def\instQU{$^{20}$}		
\def\instUSD{$^{21}$}	
\def\instUCBNE{$^{22}$}	

\title{The \MJ\ Project}

\author{S.R.~Elliott\instLANL, 
C.E.~Aalseth\instPNNL,
M.~Akashi-Ronquest\instUNC$^,$\instTUNL,
M.~Amman\instLBNLed,
J.F.~Amsbaugh\instUW,
F.T.~Avignone~III\instORNL$^,$\instUSC,
H.O.~Back\instNCSU$^,$\instTUNL,
C.~Baktash\instORNL,
A.S.~Barabash\instITEP,
P.~Barbeau\instUC,
J.R.~Beene\instORNL,
M.~Bergevin\instLBNLnsd,
F.E.~Bertrand\instORNL,
M.~Boswell\instUNC$^,$\instTUNL,
V.~Brudanin\instJINR,
W.~Bugg\instUT,
T.H.~Burritt\instUW,
Y-D.~Chan\instLBNLnsd,
T.V.~Cianciolo\instORNL, 
J.~Collar\instUC,
R.~Creswick\instUSC,
M.~Cromaz\instLBNLnsd,
J.A.~Detwiler\instLBNLnsd,
P.J.~Doe\instUW, 
J.A.~Dunmore\instUW,
Yu.~Efremenko\instUT,
V.~Egorov\instJINR,
H.~Ejiri\instOU,
J.~Ely\instPNNL,
J.~Esterline\instDUKE$^,$\instTUNL,
H.~Farach\instUSC, 
T.~Farmer\instPNNL,
J.~Fast\instPNNL,
P.~Finnerty\instUNC$^,$\instTUNL,
B.~Fujikawa\instLBNLnsd,
V.M.~Gehman\instLANL,
C.~Greenberg\instUC,
V.E.~Guiseppe\instLANL,
K.~Gusey\instJINR,
A.L.~Hallin\instUA,
R.~Hazama\instOU,
R.~Henning\instUNC$^,$\instTUNL,
A.~Hime\instLANL,
T.~Hossbach\instPNNL,
E.~Hoppe\instPNNL,
M.A.~Howe\instUW,
D.~Hurley\instLBNLnsd,
B.~Hyronimus\instPNNL,
R.A.~Johnson\instUW,
M.~Keillor\instPNNL,
C.~Keller\instUSD,
J.~Kephart\instNCSU$^,$\instTUNL$^,$\instPNNL,
M.~Kidd\instDUKE$^,$\instTUNL,
O.~Kochetov\instJINR,
S.I.~Konovalov\instITEP,
R.T.~Kouzes\instPNNL,
K.T.~Lesko\instLBNLinpa$^,$\instUCB,
L.~Leviner\instNCSU$^,$\instTUNL,
P.~Luke\instLBNLed,
S.~MacMullin\instUNC$^,$\instTUNL,
M.G.~Marino\instUW,
A.B.~McDonald\instQU,
D.-M.~Mei\instUSD,
H.S.~Miley\instPNNL,
A.W.~Myers\instUW,
M.~Nomachi\instOU,
B.~Odom\instUC,
J.~Orrell\instPNNL,
A.W.P.~Poon\instLBNLnsd,
G.Prior\instLBNLnsd,
 D.C.~Radford\instORNL,
J.H.~Reeves\instPNNL,
K.~Rielage\instLANL,
N.~Riley\instUC,
R.G.H.~Robertson\instUW, 
L.~Rodriguez\instLANL,
K.P.~Rykaczewski\instORNL,
A.G.~Schubert\instUW,
T.~Shima\instOU,
M.~Shirchenko\instJINR,
V.~Timkin\instJINR,
R.~Thompson\instPNNL,
W.~Tornow\instDUKE$^,$\instTUNL,
C.~Tull\instLBNLnsd,
T. D.~Van Wechel\instUW,
I.~Vanyushin\instITEP,
R.L.~Varner\instORNL,
K.~Vetter\instLBNLnsd$^,$\instUCBNE,
R.~Warner\instPNNL,
J.F.~Wilkerson\instUW
J.M.~Wouters\instLANL,
E.~Yakushev\instJINR,
A.R.~Young\instNCSU$^,$\instTUNL,
C.-H.~Yu\instORNL,
V.~Yumatov\instITEP,
Z.-B.~Yin\instUSD,
}

\address{ \instLANL\ Physics Division, Los Alamos National Laboratory, Los Alamos, NM, USA}
\address{\instPNNL\ Pacific Northwest National Laboratory, Richland, WA, USA}
\address{\instUNC\ Dept. of Physics and Astron., University of North Carolina, Chapel Hill, NC, USA}
\address{\instTUNL\ Triangle Universities Nuclear Laboratory, Durham, NC, USA}
\address{\instLBNLed\ Engineering Division, Lawrence Berkeley National Laboratory, Berkeley, CA, USA}
\address{\instUW\ Center for Nucl. Phys. and Astrophys., Univ. of Washington, Seattle, WA, USA}
\address{\instORNL\ Oak Ridge National Laboratory, Oak Ridge, TN, USA}
\address{\instUSC\ Dept. of Physics and Astronomy, University of South Carolina, Columbia, SC, USA}
\address{\instNCSU\ Department of Physics, North Carolina State University, Raleigh, NC, USA}
\address{\instITEP\ Institute for Theoretical and Experimental Physics, Moscow, Russia}
\address{\instUC\ University of Chicago, Chicago, IL, USA}
\address{\instLBNLnsd\ Nuclear Science Division, Lawrence Berkeley National Lab., Berkeley, CA, USA}
\address{\instJINR\ Joint Institute for Nuclear Research, Dubna, Russia}
\address{\instUT\ Dept. of Physics and Astronomy, University of Tennessee, Knoxville, TN, USA}
\address{\instOU\ Res. Center for Nucl. Phys. \& Dept. of Phys., Osaka Univ., Ibaraki, Osaka, Japan}
\address{\instDUKE\ Department of Physics, Duke University, Durham, NC, USA}
\address{\instUA\ Centre for Particle Physics, University of Alberta, Edmonton, Alberta, Canada}
\address{\instLBNLinpa\ Inst. for Nucl. \& Part. Astro., Lawrence Berkeley Natl. Lab., Berkeley, CA, USA}
\address{\instUCB\ Physics Department, UC Berkeley, Berkeley, CA, USA}
\address{\instQU\ Dept. of Physics, Queen's University at Kingston, Kingston, Ontario, Canada}
\address{\instUSD\ Dept. of Earth Science and Phys., Univ. of South Dakota, Vermillion, SD, USA}
\address{\instUCBNE\ Department of Nuclear Engineering, UC Berkeley, Berkeley, CA, USA}
\date{\today}

\ead{elliotts@lanl.gov}

\begin{abstract}
Building a \BBz\ experiment with the ability to probe neutrino mass in the inverted
hierarchy region requires the combination of
a large detector mass sensitive to \BBz, on the order of 1-tonne,
and unprecedented background levels, on the order of or less than 1 count per year
in the \BBz\ signal region.
The \MJ\ Collaboration proposes a design based on using
high-purity enriched \nuc{76}{Ge} crystals deployed in ultra-low background
electroformed Cu cryostats and using modern analysis techniques that
should be capable of reaching the required sensitivity while also being scalable to a 1-tonne size.
To demonstrate feasibility,  the collaboration plans to construct a
prototype system, the \MJ\ \DEM, consisting of 
30~kg of 86\% enriched \nuc{76}{Ge} detectors and 30~kg 
of natural or isotope-76-depleted Ge detectors.  We plan to deploy and evaluate
two different Ge detector technologies, one based on a p-type configuration
and the other on n-type.  
\end{abstract}

\section{Introduction}

This is an exciting time in our quest to understand
neutrinos --- fundamental particles that play key roles in the early
universe, in cosmology and astrophysics, and in nuclear and particle
physics. Recent results from atmospheric, solar, and reactor-based
neutrino oscillation experiments (Super-Kamiokande, SNO, and
KamLAND)\cite{Ashie:2004, Ahm04, Ara04}
have provided compelling evidence that neutrinos have mass
and give the first indication after nearly forty years of study that
the Standard Model (SM) of nuclear and particle physics is incomplete.

With the realization that neutrinos are massive, there is an increased
interest in investigating their intrinsic properties.  Understanding the neutrino nature (Majorana or Dirac),
the neutrino mass generation mechanism, the absolute neutrino mass
scale and the neutrino mass spectrum are some of the main focii
of future neutrino experiments.

Lepton number, $L$, is 
conserved in the Standard Model because neutrinos are assumed to be massless and there
is no chirally right-handed neutrino field.  The guiding principles for extending the
Standard Model are the conservation of electroweak isospin and renormalizability,
which do not preclude each neutrino mass eigenstate $\nu_i$ to be identical to
its anti-particle $\bar{\nu}_i$, or a ``Majorana" particle.   However, $L$ is no longer
conserved if $\nu=\bar{\nu}$.   Theoretical models, such as the seesaw mechanism
that can explain the smallness of neutrino mass, favor this scenario.
Therefore, the discovery of Majorana neutrinos would have profound
theoretical implications in the formation of a new Standard Model
while possibly yielding insights into the origin of mass itself.    If
neutrinos are Majorana particles, they may fit into the leptogenesis
scenario for creating the baryon asymmetry, and hence ordinary
matter, of the universe.  As of yet, there is no firm experimental
evidence to confirm or refute this theoretical prejudice.

Many even-even nuclei are forbidden to undergo $\beta$ decay, but are unstable with respect to the
second order weak process of two-neutrino double beta decay (\BBt). In this process, the nucleus
emits 2 $\beta$ particles and 2 $\bar{\nu}$. This is an allowed process and has been observed.
A similar process, neutrinoless double-beta decay (\BBz) can occur if a neutrino is exchanged between
two neutrons and no neutrinos are emitted. In this case, the only leptons in the final state are the two electrons and hence the
decay violates $L$ by two units. This exchange of the neutrino however, will only occur if the neutrino is a 
massive Majorana particle and therefore
experimental evidence of \BBz\
would establish the Majorana nature of neutrinos. The science of \BB\ has been described in detail in
many recent reviews~\cite{ell02,ell04,bar04b,eji05,avi05,avi08}.

The most-restrictive upper limits on the \BBz\ half-life come from Ge detector experiments~\cite{bau99,aal02a}. The  half-life is $>$1.9$\times$10$^{25}$ y and this corresponds to an effective Majorana mass limit from \BBz\  greater than about 400 meV depending on the choice of nuclear matrix element. 
The oscillation experiments, however, indicate that at least one of the neutrino mass eigenvalues is greater than about 
45~meV. In the inverted hierarchy, this would imply an effective Majorana neutrino mass 45~meV or greater.
The predicted half-life for \BBz\ with an effective mass in this region is greater than 10$^{27}$ years. An experiment will
require approximately 1 tonne of isotope
to be sensitive to such a long half-life, and it will require a background level of 1 count/tonne-year or better.

\section{The \MJ\ \DEM\  }
The objective of the first experimental phase of \MJ\
is to build a 60-kg module of high-purity Ge, of which 30 kg will be enriched to 86\% in $^{76}$Ge,
to search \BBz.  This module is referred to by the collaboration as the \DEM. The physics goals 
for this first phase are to:
\begin{enumerate}
\item Probe the neutrino mass region above 100~meV
\item Demonstrate that backgrounds at or below
1~count/tonne-year in the \BBz-decay region of interest 
can be achieved that would justify scaling up to a 1~tonne or larger
mass detector. 
\item Definitively test the recent claim\cite{kla06}  
of an observation of \BBz\ decay in $^{76}$Ge in the mass region around
400~meV.
\end{enumerate}


\begin{figure}
\vspace{9pt}
\begin{center}
\includegraphics[width=7cm]{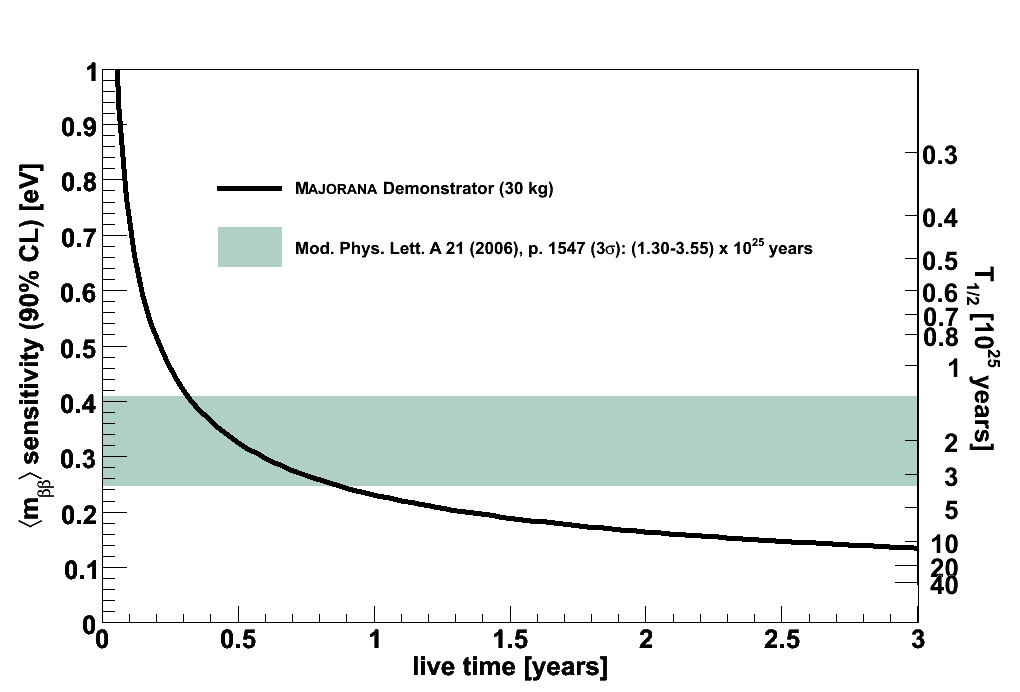}
\includegraphics[width=7cm]{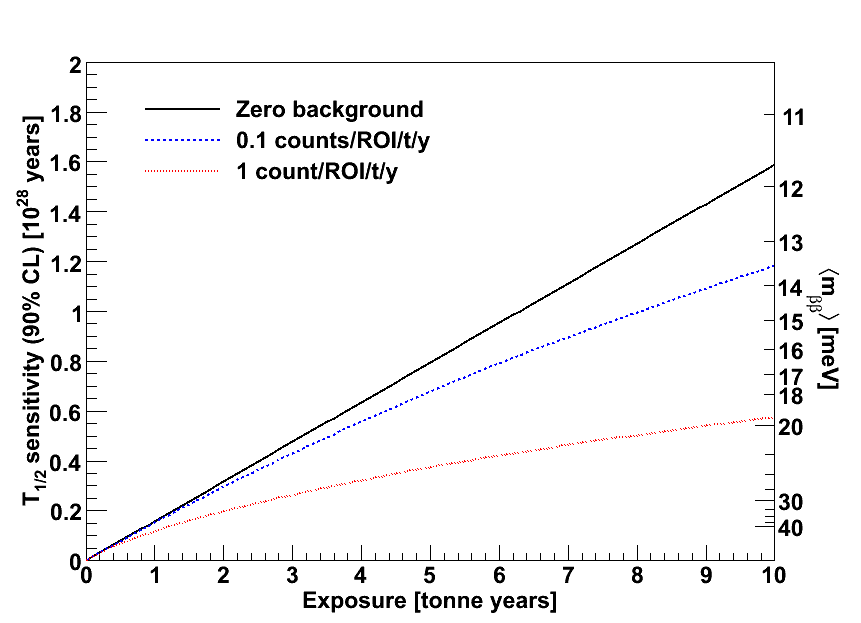}
\end{center}
\caption{The left figure shows the sensitivity of the \MJ\  \DEM\  module as a function of time and indicates the range of values indicated by a recent controversial claim\cite{kla06}. The right figure shows 
the sensitivity of a 1-ton experiment as a function of exposure for 3 different background levels. Note the factor of 1000 difference in the scales of the two panels. The matrix elements used are from Ref.~\cite{rod06}}
\label{fig:SensitivityExposure}
\end{figure}

The half live of \BBz\ is at least 10$^{25}$ years, making it an extremely difficult process to observe and therefore a very sensitive experiment is required. The \MJ\ \DEM\ will
consist of $^{76}$Ge  detectors, deployed in multi-crystal modules,
located deep underground within a low-background shielding environment. The technique 
will be augmented with recent improvements in
signal processing and detector design, and advances in controlling
intrinsic and external backgrounds.  
The justification for a detector mass size of 60 kg is directly
linked to all three of the above-stated science goals. The enriched
Ge is handled differently than natural Ge and the isotopic makeups are different. Since both these
issues lead to background differences, enriched material must also be used to
validate the background model.

\begin{figure}[htbp]
\begin{center}
\includegraphics[width=5 cm]{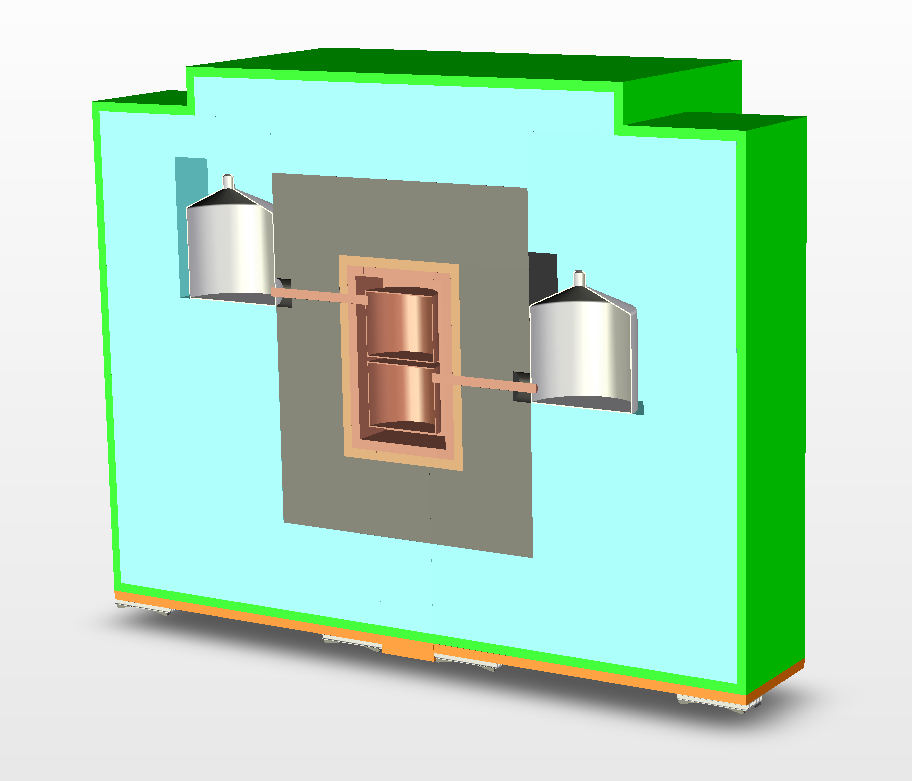}
\includegraphics[width=4 cm]{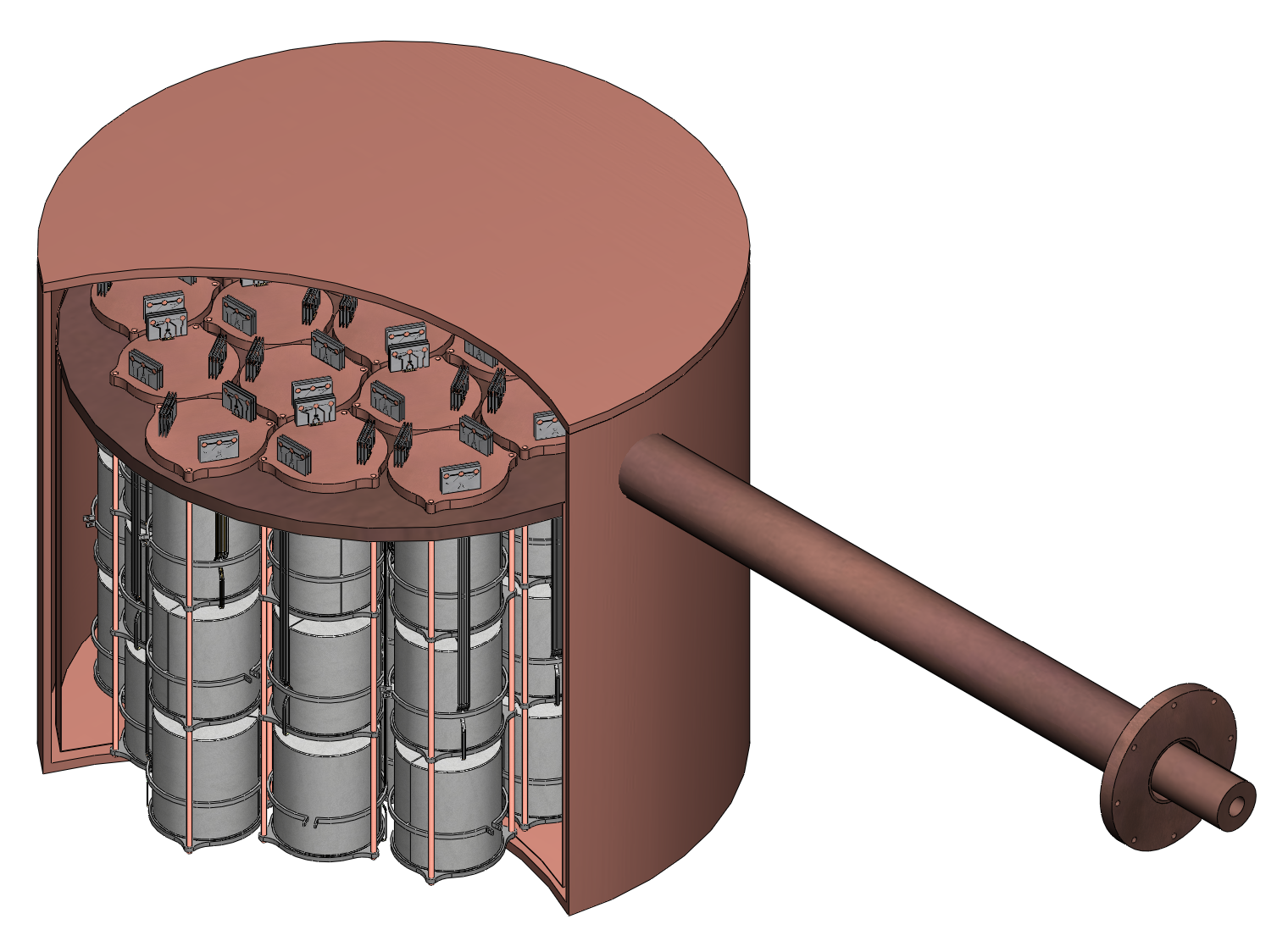}
\includegraphics[width=4 cm]{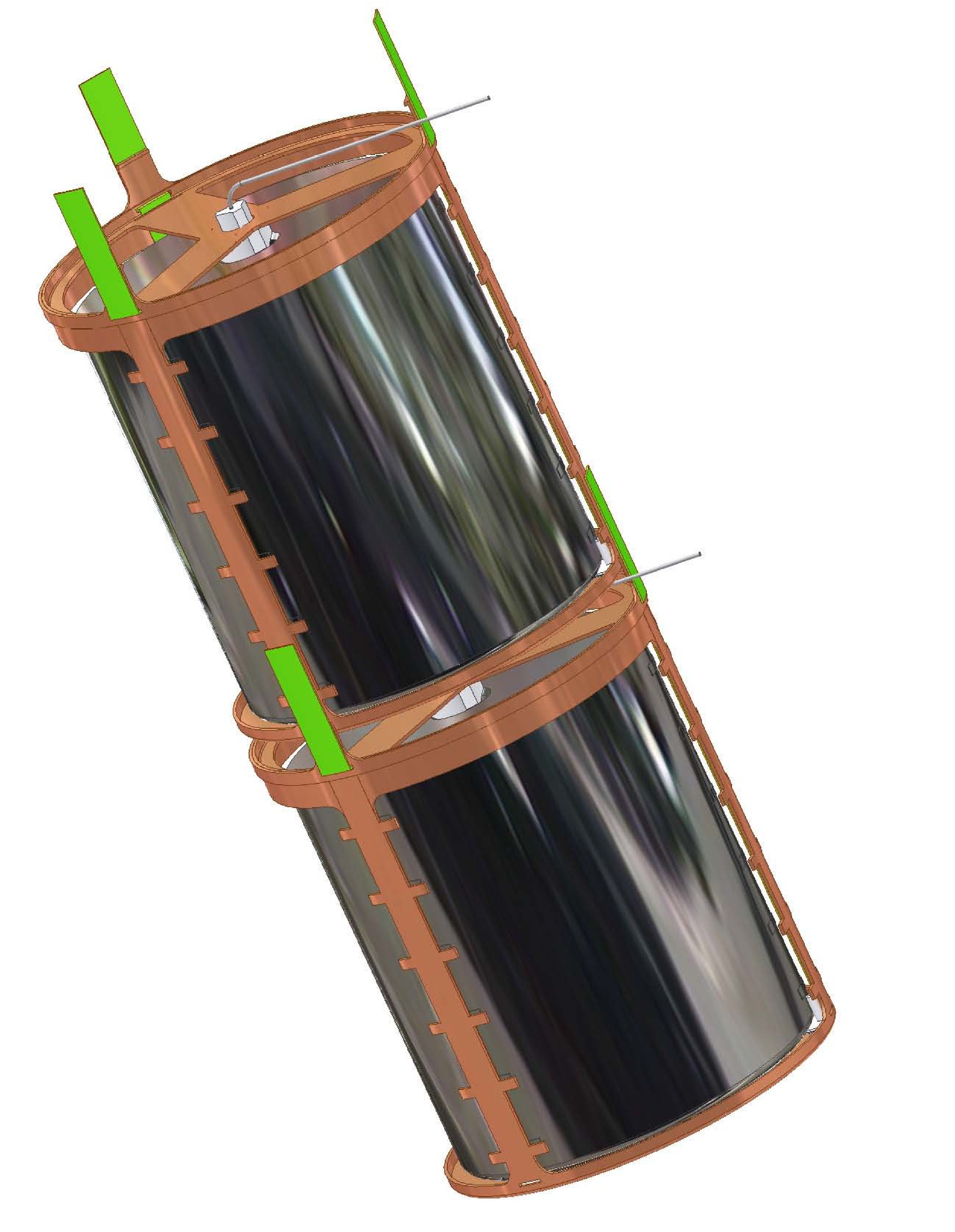}
\caption{Left: Setup of proposed \MJ\ \DEM. The 
cryostats will be built of ultrapure electroformed copper.
The inner passive shield will also be constructed of electroformed copper
surrounded by lead, which itself is surrounded by an
active muon veto and neutron moderator. Middle: A conceptual design of a 3x19 detector arrangement of Ge detectors mounted in 
one cryostat, envisioned as a building block for a tonne-scale experiment. Right: A conceptual design of a two \nsc\ detector string.}
\label{fig:MJexperiment}
\end{center}
\end{figure}

The goal of the proposed \MJ\  \DEM\ is the construction of an instrument
that provides sufficient sensitivity to test the recent
claim, that allows
a path forward to achieving a background level below 1 count/tonne-year
in the 4-keV region of interest, and is scalable towards a large-scale
instrument. We plan a module consisting of three cryostats
each containing about 20~kg of HPGe detectors. One module will
consist of 14 n-type segmented-contact (\nsc) HPGe detectors, while the
other two modules
will each consist of 28 smaller p-type point-contact (\ppc) HPGe detectors
without segmentation. Approximately 40 of the \ppc\ detectors will be
built of $\sim$30~kg of enriched $^{76}$Ge material.
Employing these two complementary detector configurations allows us
to evaluate and compare the two most promising implementations under
the realistic conditions of an ultra-low background experiment. The
multi-cryostat approach allows us to optimize each individual implementation,
providing a fast deployment with minimum interference with already
operational detectors. In addition, it allows us to separate both
configurations in the data analysis.  The proposed configuration is illustrated
in Figure~\ref{fig:MJexperiment}. Our initial emphasis will be on the first cryostat
assembled with natural-Ge \ppc\ detectors.

The efficient commercial production of Ge detectors depends strongly
on the yields for producing high-purity single crystals of Ge and
for producing p-n junction diodes from these crystals.  Single-crystal
boules naturally favor right circular cylindrical geometries, making
coaxial detectors the natural choice to efficiently use the precious
enriched Ge material.  To minimize the passivated surface area and
maximize the sensitive volume, coaxial detectors are typically
fashioned in so-called closed-end geometries with a bore hole
extending $\sim$80\% along the detector axis.  This bore hole, with
a central electrode, is necessary to produce sufficiently high
electric fields to achieve not only depletion throughout the
crystal but also high drift velocity of the charge carriers to minimize
charge trapping effects.

Previous-generation \BBz\ decay searches in $^{76}$Ge favored
detectors made of p-type material primarily because p-type crystals
can be grown more efficiently to larger dimensions than n-type
crystals.  In addition, with p-type detectors trapping effects are
reduced, and generally slightly better energy resolution is obtained.
P-type detectors are typically fabricated with a p+ B-implanted
contact on the inner bore hole and an n+ Li-diffused contact on the
outside electrode to obtain efficient and full depletion from the
outside.  While the inside B contact is very thin (typically
$<$1~$\mu$m) the Li-diffused contact is typically more then 100~$\mu$m
thick and can be up to 1 mm or more depending on the fabrication process and any subsequent annealing or prolonged storage at room temperatures. The thickness of the outside-Li contact provides an advantage
 since this dead layer absorbs alpha-particle radiation
from surface contamination that may otherwise generate background
for $^{76}$Ge \BBz\ decay.

While the pulse-shape obtained at the central contact in coaxial
detectors\cite{aal98} can provide radial separation of multiple interactions
in any implementation, pulse-shapes obtained in segmented detectors\cite{ell06a, abt06}
provide improved sensitivity in the radial separation and, more
importantly, in complementary directions as well. A high-degree of
segmentation, such as a 6x6-fold segmentation, enables the full
reconstruction of $\gamma$-ray interactions within the detectors.
GRETINA in particular has provided very encouraging analyses that
events can be reconstructed with a position uncertainty of $\sim$2~mm~\cite{vet00}.
Even without absolute event vertex reconstruction, events with
multiple energy depositions can be identified and rejected.
Preliminary results from GRETINA and other highly-segmented HPGe
arrays indicate that a minimum separation of 4~mm will be achievable.
However, these advanced capabilities come at the price of a
proportionally larger number of small parts such as cables near the detector, and
their selection must be optimized against their ability to overcome
added backgrounds.

\begin{figure}[t]
\begin{center}
\includegraphics[width=10 cm]{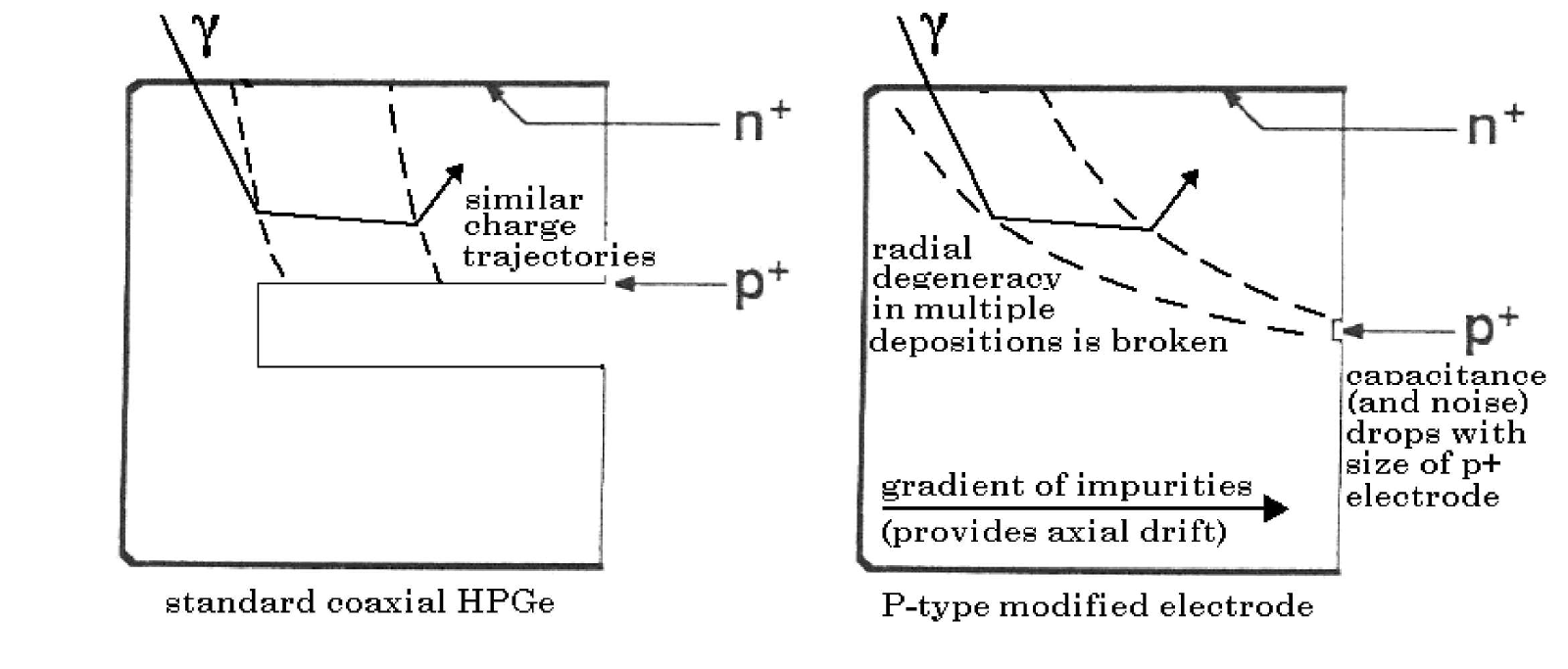}
\caption{Closed-end coaxial p-type Ge detector (left) versus 
the p-type point contact Ge detector implementation (right). Figure is adapted from Ref.~\cite{barb06}}
\label{fig:pmeopt}
\end{center}
\end{figure}

An alternative right circular cylindrical detector design has been
developed in which the bore hole is removed and in its place a
point-contact, either B implanted  or Li diffused, is formed at the center
 of the intrinsic (open end) detector surface ~\cite{luk89}.  The
changes in the electrode structure result in a drop in capacitance
to $\sim$1~pF, reducing the electronics noise component and enabling
sub-keV energy thresholds.  This point-contact 
configuration also has lower electric fields throughout the bulk
of the crystal and a weighting potential that is sharply peaked
near the point contact. (Earlier attempts employing n-type material exhibited poor energy
resolution at hight energies due to increased charge trapping.)
This in turn results in an increased range
of drift times and a distinct electric signal marking the arrival
of the charge cloud at the central electrode.  Figure~\ref{fig:pmeopt}
illustrates the p-type point-contact (\ppc) implementation in contrast to the
conventional, closed-end coaxial detector approach.  Instrumented
with a modern FET, a \ppc\ detector was recently demonstrated by
\MJ\ collaborators to provide low noise, resulting in a low energy
trigger threshold and excellent energy resolution, as well as
excellent pulse-shape capabilities to distinguish multiple
interactions~\cite{barb06}.  This detector was developed with the
goal of detecting the very soft (sub-keV) recoils expected from
coherent neutrino-nucleus scattering in a reactor experiment.

The \ppc\ detector in~\cite{barb06} has a diameter of 5~cm and a
length of 5~cm resulting in a total mass of 0.5~kg. While it may
be possible to expand the radial dimension of these detectors, it will be more difficult to expand the
axial dimension due to the
requirement of a significant impurity gradient between the detector
faces.  Without an impurity concentration gradient, the electric
fields can be sufficiently weak that the detector is susceptible
to excessive charge trapping, even in the p-type configuration. It is
interesting to note that increasing the voltage, {\em e.g.}~beyond full depletion,
will not significantly increase the electric field in the detector
bulk away from the point contact.  In contrast, the presence of
space charge due to an appropriate impurity concentration and gradient
 is able to increase the electric field throughout the
crystal.  The first attempt to make a \ppc\ detector with a flat
impurity concentration failed.  The detector in~\cite{barb06} resulted
from a second attempt with a factor two difference in impurity
concentration between the detector faces. The collaboration 
has recently purchased and received several additional \ppc\ detectors from various vendors.

Although the detector mounts and feedthroughs require different
designs, the overall dimensions of the cryostats can be similar,
reducing the design effort and simplifying production. Fabricating
multiple, similar cryostats will provide an opportunity to assess several risks associated with a large-scale
modular deployment for a 1-tonne experiment. These risks include
the reproducibility of the mechanical and thermal integrity of the
cryostats, as well as the ability to consistently achieve and
maintain radiopurity throughout the module production and assembly
chain.  Figure~\ref{fig:MJexperiment} shows the conceptual design of a
larger, 57-detector cryostat currently envisioned for a tonne-scale
modular deployment of $^{76}$Ge detectors.

Following the selection of two different detector designs, {\em i.e.}~the
\nsc\ and \ppc\ configurations, the mechanical components within
the cryostat will be optimized independently. However, the overall
dimensions and wall thicknesses can be the same since each cryostat
will contain about 20~kg of Ge detectors.  Our reference design is
a cylindrical vessel with a thick cold plate at the
top from which the detectors hang.  A thermal shroud mounted to the
cold plate provides radiative cooling for the detector array.
The cryostat, cold plate and thermal shroud will be fabricated using
ultra-low-back\-ground, relatively-thick electroformed copper
to eliminate any concern about collapse of the
vacuum vessel.  The coldplate shields the germanium detectors from
backgrounds originating in the front-end electronics; thus, a thicker
coldplate is favored.  The cryostats themselves are situated within a
thick layer of electroformed Cu comprising the innermost
passive shielding layers, so extra cryostat wall and coldplate
thickness is not a concern from the standpoint of backgrounds.
However, thicker cryostat walls do reduce the potential for
inter-cryostat granularity rejection.  A combination of thermal,
mechanical and background analyses will be used to determine the
preferred thickness of these components for the cryostats in the
\DEM\ with the upper limit being set by considerations
of the copper electroforming process (time and material quality).

The detector mounts for each configuration will be different, as
will the number of cables and feedthroughs.  While the assembly and
readout for the central channel in the \nsc\ configuration will be
similar to the \ppc\ approach, all electronics components for the
segment signals will be located outside the cryostat, at the maximum
distance possible to minimize radioactive background in the detectors.
This distance is constrained by the bandwidth and noise requirements
for the pulse-shape analysis to achieve sufficent position sensitivity.
Using warm FETs for the segments also reduces the heat load to the
cryostat.  Nevertheless, beyond the feedthroughs for high voltage,
test input, feedback, ground, and signal for the central channel
for each detector, \nsc\ detectors require an additional feedthrough
for each outer segment of each crystal, resulting in an additional thermal
burden to the cryostat and background burden to the detectors.  Assuming, for comparison,
 28 750-g \ppc\ detectors arranged
in 4 layers of 7 detectors for each of the two cryostats, each
\ppc\ cryostat will have 140 feedthroughs.  In contrast, a cryostat with 14 1.5-kg \nsc\
36-segment detectors arranged in 2 layers of 7 detectors will
have 574 feedthroughs.  

In both configurations, the detectors will
be mounted in a string-like arrangment as shown in Figure~\ref{fig:MJexperiment}
for the \nsc\ configuration. This reference design consists of a
thick copper ``lid'', a copper support frame, and low-mass Teflon
support trays or standoffs.  Each individual detector is mounted
in a separate frame built of electroformed Cu that eases handling
of individual detectors during shipping, acceptance testing and
string assembly.  The assembled detector string is simply lowered
through a hole in the cryostat cold plate until the string lid sits
on the cold plate.  This allows us to easily mount and dismount
individual detector strings from the top. Cables are run vertically
from the detector contacts through a slit in the string lid, above
which low-background front-end electronics packages are mounted to read
out the central contact. For \nsc\ detectors, HV blocking
capacitors will also be mounted above the lid.  The string lid
provides some shielding between these components and the detectors,
and allows the bandwidth to be maximized by placing the central
contact FETs near the germanium detectors.  The front-end and the
outer contact leads are routed out of the detector along the cold
finger.

The \MJ\ \DEM\ cryostats will be enclosed in a graded passive shield and
an active muon veto to eliminate external backgrounds.
Shielding reduces signals from
$\gamma$ rays originating in the experiment hall, cosmic-ray $\mu$'s penetrating the shielding, and
cosmic-ray $\mu$-induced neutrons. The strategy is to provide
extremely low-activity material for the  inner shield. Surrounding
this will be an outer shield of bulk $\gamma$-ray shielding material 
with lower radiopurity.  This high-$Z$ shielding
enclosure will be contained inside a gas-tight Rn exclusion box
made of stainless-steel sheet. Outside this bulk high-$Z$ shielding
will be a layer of hydrogenous material, some of which will be doped
with a neutron absorber such as boron, intended to reduce the neutron
flux.  Finally, active cosmic-ray anti-coincidence detectors
will enclose the entire shield. The \MJ\ collaboration plans to site the \DEM\  deep 
underground in the Sanford Laboratory at the 
Homestake gold mine in Lead, South Dakota, USA. 

The data acquisition software system will be constructed using the Object-oriented
Real-time Control and Acquisition (ORCA)~\cite{How04} application to achieve the
goal of providing a general purpose, highly modular, object-oriented,
acquisition and control system that is easy to develop, use, and
maintain. The object-oriented nature of ORCA enables a user to
configure it at run-time to represent different hardware configurations
and data read-out schemes by dragging items from a catalog of objects
into a configuration window. 

Monte Carlo (MC) radiation transport simulation models have been
developed for \MJ\ using MaGe~\cite{cha08}, an object-oriented MC simulation
package based on ROOT~\cite{Bru97} 
and the Geant4~\cite{Ago03, Alli06} toolkit and 
optimized for simulations
of low-background germanium detector arrays. MaGe is being jointly
developed by the \MJ\ and GERDA~\cite{Abt04} collaborations in consultation with collaborators from the
National Energy Research Scientific Computing Center (NERSC) at
LBNL. MaGe defines a set of physics processes, materials, constants,
event generators, etc.~that are common to these experiments, and
provides a unified framework for geometrical definitions, database
access, user interfaces, and simulation output schemes in an effort
to reduce repetition of labor and increase code scrutiny.

\section{Conclusion}
The \MJ\ collaboration is carrying out R\&D to develop a 1-tonne \BBz\ experiment based on enriched $^{76}$Ge. The Ge detectors will be housed in a classic cryostat design that is constructed from ultra low-radioactivity materials. The cryostats will be contained within a passive shield, and active veto system and operated deep underground. The project should demonstrate that this technology will be low enough in background to verify the feasibility of constructing a 1-tonne Ge-based \BBz\ experiment.

\section*{Acknowledgments}
This conference was organized to celebrate the 75$^{th}$ birthdays of Frank Avignone, Etorre Fiorini and Peter Rosen. The entire field of \BB\ research has benefitted from the long and storied efforts of Frank and Etorre. I admire their scientific leadership at this time we celebrate their careers. The contributions to this field by Peter Rosen are also paramount. It is with fondness that we remember his legacy, but with sadness that we can't also share his 75$^{th}$ birthday at this event.

The author wishes to acknowledge the contributions of the \MJ\ Collaboration. Our efforts have also benefitted from a close cooperation with the GERDA collaboration. The preparation of this manuscript was supported
in part by Laboratory Directed Research and Development at LANL.

\section*{References}

\bibliographystyle{iopart-num}
\bibliography{DoubleBetaDecay.bbl}

\end{document}